\documentclass[10pt]{IEEEtran}

\usepackage{graphicx}
\usepackage{listings}
\usepackage{url}
\usepackage[top=1.0in]{geometry}
\begin{document}


%
\title{The Program with a Personality:\\ Analysis of Elk Cloner, the First Personal Computer Virus}
\date{}



%
\author{\IEEEauthorblockN{Scott Levy\IEEEauthorrefmark{1} and
Jedidiah R. Crandall\IEEEauthorrefmark{2}\IEEEauthorrefmark{3}\\}
\IEEEauthorblockA{\IEEEauthorrefmark{1}Sandia National
Laboratories\\sllevy@sandia.gov\\}
\IEEEauthorblockA{\IEEEauthorrefmark{2}Arizona State
University\\jedimaestro@asu.edu\\}
\IEEEauthorblockA{\IEEEauthorrefmark{3}Breakpointing Bad\\breakpointingbad.com\\}
}

\maketitle

\begin{abstract}


Although self-replicating programs and viruses have existed since the 1960s and
70s, Elk Cloner was the first virus to circulate among personal computers in
the wild.  Despite its historical significance, it received comparatively
little attention when it first appeared in 1982.  In this paper, we: present
the first detailed examination of the operation and structure of Elk Cloner;
discuss the effect of environmental characteristics on its virulence; and
provide supporting evidence for several hypotheses about why its release was
largely ignored in the early 1980s.
\end{abstract}


%

\section{Introduction}

Self-replicating programs have existed since John von Neumann's work with
cellular automata in the 1960's.  Programs built on the idea of
self-replication began to appear in the form of computer viruses on mainframe
computers in the 1970's.\footnote{Frederick Cohen did not actually coin the
term ``computer virus'' until 1984.  \cite{cohen}  However, given that many
earlier self-replicating programs exhibited virus-like behavior, they have
retroactively been given the label of ``virus''.}  Creeper appeared on DEC
PDP-10s in the early 1970s and PERVADE emerged in 1975 on a UNIVAC 1108.  These
early viruses were confined largely (if not exclusively) to business and
research machines; it was another decade before they made the leap to personal
microcomputers.  \cite{Szor}\\
\\
In 1982, Rich Skrenta (then a high-school student) released the first virus
known to circulate in personal microcomputers: Elk Cloner.  He began
experimenting with self-replicating programs as a way to play practical jokes
on his classmates. \cite{Skrenta} Elk Cloner infected Apple ][ Plus diskettes
and was largely benign.\footnote{Some diskettes infected with Elk Cloner crash
unexpectedly or fail to boot properly.}  It exhibited many behaviors but was
best known for printing a poem on the fiftieth boot of an infected diskette.
\\
\begin{figure}[htb]
\begin{center}
\includegraphics[width=\columnwidth]{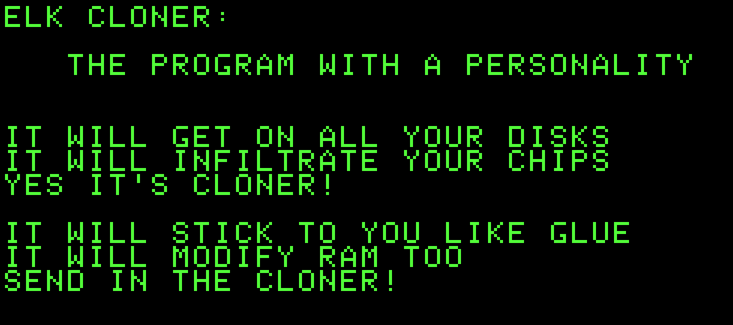}
\end{center}
\caption{The poem displayed by Elk Cloner on the fiftieth boot of an infected diskette.}
\end{figure}
{\ \\}
In the intervening decades since Elk Cloner was first released, viruses have
multiplied, evolved complex behaviors and spawned a multi-billion dollar
industry.  Given the influence that threats to computer security have on modern
computing, it is instructive to understand how personal computer viruses
emerged.  In this article, we present the first detailed description of the
operation of Elk Cloner: how it loaded itself into memory, how it spread from
disk to disk and how it made its presence known to users.  


\section{Overview of Apple ][ and Apple DOS}\label{sec:overview}
The first Apple ][ appeared in June 1977.  It shipped with 4 kilobytes of
memory (upgradeable to 48 kilobytes) and a NMOS 6502 processor.  Two years
later, the Apple ][ Plus was released.  It offered a minor upgrade over the
original.  The smaller memory configurations of the Apple ][ were eliminated;
the Apple ][ Plus shipped with no less than 48 kilobytes of memory (upgradeable
to 64 kilobytes with the addition of a language card).  Additionally, a ROM
containing Applesoft BASIC was included on the Apple ][ Plus.\\
\\
The 6502 was an 8-bit microprocessor running at 1 MHz.  It was capable of
executing 56 different instructions, used 9 addressing modes and provided only
three registers for general purpose use: the Accumulator (A), an X register and
a Y register.  \cite{MOSprogrammers}  Although the 6502 was an 8-bit processor,
it used a 16-bit address space.  It accomplished this by using two 8-bit
registers (\texttt{ADH} and \texttt{ADL}) to store addresses.  The upper 8-bits
were stored in the \texttt{ADH} register and represented the page to be
accessed and the lower 8-bits were stored in the \texttt{ADL} register and
contained the offset within the page.  A similar arrangement was used to store the program
counter.  \cite{MOShardware}
\\
Unlike modern computers, the Apple ][ Plus had no integrated persistent storage
(e.g., a hard drive) and no operating system in the modern sense of the word.
It provided no virtual memory or multitasking support.  As a result, every
application was single-threaded and applications were typically loaded from
diskettes.\footnote{Programs and data could also be loaded from cassette tape.}
Diskettes frequently included a copy of Apple DOS:\footnote{Alternatively, some
programs used ProDOS instead of Apple DOS.  However, as we will see, Elk Cloner
relied upon the structure of Apple DOS 3.3 to replicate itself.  Moreover, a
ProDOS disk infected with Elk Cloner was inoperative. Therefore, we will
confine our discussion to Apple DOS.} a set of routines that simplified
hardware access.

%
\subsection{Boot Process}
When an Apple ][ booted an Apple DOS diskette, it began executing a sequence of
successively larger and more complex boot loaders beginning with ``Boot 0''
which resided on a ROM.  Once in memory, Boot 0 loaded ``Boot 1'' from Track 0,
Sector 0 of the diskette into memory.  Boot 1 then loaded the final boot stage,
``Boot 2'' into memory from Track 0, Sectors 1 through 9 of the diskette.  Boot
2 then loaded DOS and the Relocator into memory.  A DOS diskette could be
either a ``master'' disk or a ``slave'' disk.  The primary distinction between
the two was that a master diskette could be booted on any Apple ][ regardless
of its memory configuration.  This flexibility relied on the Relocator.  Boot 2
would load a copy of DOS into an address that was valid on even the smallest
memory configurations and the Relocator would then move the DOS image based on
the actual amount of physical memory available.  On the other hand, a slave
diskette loaded DOS into a fixed address in memory so no Relocator was
required.  \cite{BeneathDOS}
%
\subsection{Diskette Organization}
Apple ][ diskettes had 35 tracks.  Starting with DOS 3.3 (released in August
1980), each track consisted of 16 256-byte sectors.  On an Apple DOS diskette, the
first three tracks contained the boot loaders and a copy of DOS.  Track 17,
Sector 0 contained a Volume Table of Contents (VTOC).  The VTOC was the root of
the diskette's metadata structures.  It described the characteristics of the
diskette (e.g., the release of DOS that was used to initialize it, its volume
number and bit maps of the free sectors in each track of the diskette).  It
also was the head of the linked list of catalog sectors that allowed individual
files to be located.  Although an entire sector was reserved for the VTOC, on a
standard 35 track diskette more than a third of the sector was unused.  As we
will see, these unused bytes provided Elk Cloner with a persistent store that
was invisible to the user.\\

\section{Elk Cloner}\label{sec:elkcloner}
Elk Cloner was written in 1982 on an Apple ][ Plus.  The 6502 machine-language
source code and a disk image for Elk Cloner 2.0 have been preserved by the
virus's author and are available for download on his website.  \cite{Cloner}
Although the version number suggests otherwise, this is the virus that
circulated in the wild.  \cite{Skrenta}  By combining his intimate knowledge of
the Apple ][ with his observations about the way that his friends at school
used computers, the author of Elk Cloner was able to construct a program that
was capable of replicating itself and spreading through the diskettes of his
unsuspecting classmates.\\
\\
The Elk Cloner life-cycle consists primarily of three phases: (1) Boot Loading;
(2) Replication; and (3) Manifestation.  The Boot Loading phase is the process
by which the virus moves from an infected disk into memory.  Replication is the
process by which it infects new diskettes.  Manifestation is the process by
which the virus exhibits user-observable behavior.

\subsection{Boot Loading}
We begin by considering what happened when a diskette infected with Elk Cloner
was booted.  To understand this process, it is important to understand how an
infected diskette differed from an uninfected diskette.  A diskette that
carried an Elk Cloner infection had three important characteristics: it would
have contained (1) a copy Elk Cloner loader in one of the sectors reserved for
the Relocator; (2) a copy of DOS modified to transfer control to Elk Cloner;
and (3) a copy of the Elk Cloner executable and the Elk Cloner loader in the
unused space at the end of Track 2.\\
\\
DOS diskettes reserved Track 0, Sectors 10 and 11 for the Relocator.  On a
master diskette, these two sectors contained the code that moved DOS in memory
based on the total amount of memory available.  However, on a slave diskette no
Relocator was required and these sectors were filled with zeros.  On a diskette
infected with Elk Cloner, Track 0 Sector 10 contained the Elk Cloner loader.
As a result, a master diskette infected with Elk Cloner would not function
properly since the Relocator had been partially overwritten.\\
\\
To gain control of execution during the boot process, an infected diskette
included a modified copy of DOS.  Figure \ref{three} shows the difference
between an uninfected and an infected copy of DOS.  The modification consisted
of the replacement of the first machine instruction in the DOS common command
handler routine.  When a DOS command (e.g., \texttt{CATALOG}) was executed, the
command handler routine was the common entry point for identifying and
executing the proper subroutine.  Although the modification is simple, it was
the linchpin of Elk Cloner's infection process.\\
\\
The first three tracks of DOS diskettes were reserved for the boot loaders and
a copy of DOS.  However, the last twelve sectors (sectors 5 through 16) of
track 2 are unused (i.e., are not loaded by Boot 2).  On an infected diskette,
these sectors contain the Elk Cloner executable.  The six sectors beginning at
track 2, sector 3 contained Elk Cloner and a second copy of the Elk Cloner
loader.  Two of the sectors (Track 2, Sectors 3 and 4) used to store the Elk
Cloner executable are within the portion of the disk reserved for DOS.  Given
that Elk Cloner would fit easily in the unused sectors beginning at Track 2
Sector 5, overwriting sectors 3 and 4 seems unnecessary.  Many programs
infected with Elk Cloner appear to operate normally (after accounting for Elk
Cloner's manifestation behaviors), but it may be that some of the unexpected
crashes observed in infected diskettes are attributable to the fact that these
two sectors were overwritten.
\begin{figure}[htb]
\centering
\begin{tabular}{rll}
{Offset} & {Uninfected DOS} & {Elk Cloner Infected DOS}\\
\hline
\texttt{\$78:} & {\texttt{BEQ \$A116}} & {\texttt{BEQ \$A116}} \\
\texttt{\$7A:} & {\texttt{JSR \$A180}} & {\texttt{JSR \$A180}} \\
\texttt{\$7D:} & {\texttt{JMP \$9F83}} & {\texttt{JMP \$9F83}} \\
\texttt{\$80:} & \textbf{\texttt{JSR \$A75B}} & \textbf{\texttt{JMP \$9B00}} \\
\texttt{\$83:} & {\texttt{JSR \$A1AE}} & {\texttt{JSR \$A1AE}} \\
\texttt{\$86:} & {\texttt{LDA \$AA5F}} & {\texttt{LDA \$AA5F}} \\
\texttt{\$89:} & {\texttt{TAX}} & {\texttt{TAX}} \\
\end{tabular}
\caption{A comparison of uninfected Apple DOS (left) and Apple DOS on a diskette carrying Elk Cloner (right).  The offset is the number of bytes from the beginning of Track 1, Sector 0.  The common command handling code begins at offset \texttt{\$80}.  Other than the instructions shown here, the two instances of Apple DOS are identical.}\label{three}
\end{figure}
\\
As discussed above, when a diskette was booted, the Apple ][ began executing a
series of increasingly complex boot loaders.  Boot 2, which resided in Track 0
was responsible, in part, for loading DOS into memory.  In addition to loading
DOS from the disk, Boot 2 loaded the Relocator whether it was needed or not.
On an infected diskette, the first sector reserved for the Relocator contained
the Elk Cloner loader.  As a result, Elk Cloner could rely on Boot 2 to load
its loader into memory.\\
\\
\\
Once the Elk Cloner loader was in RAM, it was simply a matter of forcing DOS to
execute it.  After DOS was loaded, it executed the \texttt{HELLO} program by invoking the
\texttt{RUN} command handler.  At address \texttt{\$A180}\footnote{Although it
is now more common to signify hexadecimal values by preceding them by
\texttt{0x} (e.g., \texttt{0xA180}), the Apple ][ literature largely uses the
\texttt{\$} convention.  Therefore, we use this convention throughout.}  was
the DOS command handler; the common point from which all command subroutines
were launched (see Figure \ref{three}\footnote{Track 1, Sector 0 of a slave
diskette is loaded into memory beginning at \texttt{\$A100}.  Therefore, offset
\texttt{\$80} corresponds to address \texttt{\$A180}.}).  In an uninfected copy
of DOS, when a DOS command was executed by the user, this subroutine performed
some housekeeping, located the address of the subroutine that performed the
requested action, and executed it.  However, in a copy of DOS infected with Elk
Cloner, the first instruction of the command handler had been overwritten with
an unconditional branch to the Elk Cloner loader.\\
\\
At this point in the boot process, Elk Cloner had a toehold into the system;
the Elk Cloner loader was in RAM and had been given control over the execution
of the machine.  The Elk Cloner loader was a simple, small program\footnote{The
Elk Cloner loader requires only 51 bytes of machine code.} whose sole purpose
was to load the main Elk Cloner code into memory.  It copied Track 2, Sectors
3-8 into memory at \texttt{\$9000-\$9600}; we note that this includes a
duplicate copy of the Elk Cloner loader.  
Although the second copy of the loader was not strictly necessary, having a
second copy in a contiguous memory region with the main Elk Cloner code would
have simplified the cloning code (e.g., it was simply one additional iteration
of the copy loop).  Once the main Elk Cloner code was loaded, the loader
executed an unconditional branch to its beginning.\\
\\
When Elk Cloner began executing, it first wrote \texttt{\$8FFF} to the memory
at address \texttt{\$004C}.  This was the location used to indicate the highest
memory address (HIMEM) available to INTEGER BASIC.  This prevented INTEGER
BASIC programs from overwriting the memory occupied by Elk Cloner\footnote{Applesoft BASIC maintained a separate HIMEM value at a different memory
location (\texttt{\$0073}).  Elk Cloner made no attempt to modify Applesoft
BASICS's HIMEM value.}.\\
\\
The next task was to perform some housekeeping.  Elk Cloner began by repairing
the modified instruction at the beginning of the common command handling code.
At this point in the boot process, the first instruction of the command handler
code in RAM was an unconditional branch to the Elk Cloner loader.  To avoid
unnecessarily and repeatedly invoking the Elk Cloner the first instruction of
the command handler was restored to its original state (i.e., overwritten to
match an uninfected instance of DOS).  Elk Cloner also performed a similar
cleanup operation on the \texttt{RUN} command.  However, the \texttt{RUN}
command was never modified by Elk Cloner.  It may be that the virus's author
had designs on modifying the \texttt{RUN} command, but his ideas were never
fully implemented.\\
\\
Elk Cloner then proceeded to insert itself into three DOS commands:
\texttt{LOAD}, \texttt{BLOAD}\footnote{\texttt{BLOAD} loaded binary files.} and
\texttt{CATALOG}.  The DOS commands commandeered by Elk Cloner represented the
commands that a user would be most likely to enter when attempting to run a new
program from the command line.  In some circles, it was common for users to sit
down at a computer, insert a disk and launch their desired program without
rebooting.  \cite{Skrenta}  When a user first inserted a disk, the
\texttt{CATALOG} command allowed the user to see which files were on the disk.
If the user was familiar with the contents of the disk, she could skip the
\texttt{CATALOG} command and load the program into memory using either
\texttt{LOAD} (for BASIC programs) or \texttt{BLOAD} (for binary programs) and
then use \texttt{RUN} to launch the program\footnote{\texttt{RUN} could be supplied with a program name directly, but this would still be loaded with \texttt{LOAD} or \texttt{BLOAD}.}.
\\
Additionally, Elk Cloner hooked into the \texttt{USR}\footnote{User-Specified
Routine.} command.  The \texttt{USR} command was an Applesoft BASIC command
that allowed a machine-language subroutine to be invoked directly from BASIC.
The three bytes of memory starting at \texttt{\$000A} contained an instruction
that would be executed when the \texttt{USR} command was invoked.  Typically,
this memory would store an unconditional branch to a machine-language
subroutine supplied by the user.  When the user invoked the \texttt{USR}
command from an Applesoft BASIC program, the instruction at \texttt{\$000A} was
executed.  Additionally, the user could include an argument that was passed to
the machine-language program in register X of the processor.
\cite{AppleBASIC}\\
\\
When Elk Cloner executed during the boot process, it wrote an unconditional
branch to \texttt{\$000A} that pointed to Elk Cloner's \texttt{USRCMD}
subroutine.  The \texttt{USRCMD} subroutine essentially exposed an
administrative interface to Elk Cloner.  Based on the argument passed to the
\texttt{USR} command, the user could: display the version and serial number of
Elk Cloner; display the number of times that the diskette has been
booted;\footnote{The significance of the boot count is explained fully later.}
infect the current diskette; or display the famous Elk Cloner poem.\\

The Elk Cloner serial number is particularly noteworthy.  As the virus's author
began infecting diskettes with his new virus, he became curious about the way
in which it spread.  As a result, he embedded a serial number into the code
that would allow him to identify the source of the original infection.  Each
time he infected a new batch of diskettes he incremented the number so that if
he ever encountered an infected diskette, he would be able to identify which of
his releases led to the infection.  This would provide a simple metric for
evaluating the spread of the virus in the wild. \cite{Skrenta}\\
\\
Once it had inserted itself into these key commands, Elk Cloner performed one
last bit of housekeeping.  First, it loaded the VTOC of the diskette into
memory.  Elk Cloner used an unused byte in the VTOC to maintain a persistent
record of how many times the diskette has been booted.  It incremented this
count and wrote the new value back to the diskette.  Finally, before control of
execution was returned to DOS, Elk Cloner examined the current boot count to
determine whether and how it should manifest itself.  The process by which Elk
Cloner made its presence known is discussed in a later section.

\subsection{Replication}
For each of the compromised commands (\texttt{LOAD}, \texttt{BLOAD} and
\texttt{CATALOG}), Elk Cloner overwrote the first instruction of the command's
subroutine with an unconditional branch to one of Elk Cloner's command
handlers.  As a result, whenever one of these commands was invoked, control of
execution was passed to Elk Cloner.  Each command handler performed the same
three primary tasks: (1) determined whether the current diskette had already
been infected; (2) invoked Elk Cloner's replication subroutine (if necessary);
and (3) prepared for re-entry into the DOS command subroutine.\\
\subsubsection{Detecting Infections}
{\ \\}
Elk Cloner used another unused byte in the VTOC to store its version.  It used
the version information to determine whether the diskette had already been
infected.  If the version in the diskette's VTOC matched the version of the
instance of Elk Cloner that was resident in memory, then Elk Cloner concluded
that the diskette has already been infected.\\

\subsubsection{Infection Replication}
{\ \\}
If the diskette was uninfected, the command-specific block invoked Elk Cloner's
replication code. The first step of replication was to load the contents of the
diskette's VTOC into memory.  Elk Cloner wrote its version and serial number
into two unused bytes of the VTOC.  The VTOC was then written back to the
diskette.  Elk Cloner then copied its main code and a copy of the Elk Cloner
loader into Track 2, Sectors 3-8.  Next, it patched DOS so that Elk Cloner
would be invoked during the boot process.  To accomplish this task, Elk Cloner
loaded Track 1 Sector 0 into memory, overwrote the first instruction of the DOS
command handler and wrote the result back to disk.  Finally, Elk Cloner wrote a
copy of its loader into Track 1 Sector 10.  This was the region of the diskette
that was reserved for the Relocator.  As a result, Elk Cloner could rely on DOS
to transfer its loader from diskette into memory.  However, if the diskette had
already been infected, Elk Cloner skipped the replication code.

\subsubsection{DOS Re-entry}
{\ \\}
At this point, Elk Cloner had completed its work and needed to relinquish
control of execution back to DOS.  Because the first instruction of each of the
DOS command handlers (for \texttt{LOAD}, \texttt{BLOAD} and \texttt{CATALOG})
had been overwritten, Elk Cloner could not directly invoke the appropriate DOS
command handler.\footnote{An infinite loop would result, and Elk Cloner
infections would have been destructive.}  Instead, Elk Cloner executed the
instruction that it had overwritten and then transferred control to the
remaining code of the command handler.  The exception to this approach was the
code that intercepted the \texttt{CATALOG} command.  The \texttt{CATALOG}
command handler consisted of just four instructions.  As a result, Elk Cloner
included a complete copy of the DOS \texttt{CATALOG} command handler.  Elk
Cloner's \texttt{CATALOG} command handler also included what appears to be
vestigial code.  Before exiting Elk Cloner, the command handler cleared the
three bytes beginning at \texttt{\$B3BE}.  This memory contained the DOS VTOC
buffer.  The three affected bytes contained: the version of DOS that the
diskette was initialized with; the Elk Cloner boot counter; and Elk Cloner
serial number.  Given that two of these bytes are unused, clearing them seems
unnecessary.\\

\subsection{Manifestation} 
\begin{figure}[htb!]
\begin{footnotesize}
\centering
\begin{tabular}{|l|p{2in}|} \hline
{Boot \#} & {Behavior} \\
\hline
10th & {Overwrote the reset vector so that pressing {\sc control-reset} enters the Monitor program instead of DOS.} \\ \hline
15th & {Modified the video mode so that the text on the screen was inverted.} \\ \hline 
20th & {Wrote to the speaker, causing a brief click to be heard.} \\ \hline 
25th & {Modified the video mode so that the text on the screen flashed.} \\ \hline 
30th & {Rearranged the characters that represent the file type of a file when the {\tt CATALOG} command was executed} \\ \hline 
35th & {Modified the value that represented {\sc control-d} so that DOS commands invoked from Applesoft BASIC are printed instead of executed.} \\ \hline 
40th & {Overwrote the reset vector with an instruction that pointed to the reset vector so that pressing {\sc control-reset} caused the machine to enter an infinite loop.} \\ \hline 
45th & {Set the Applesoft program protection flag (e.g., attempts to {\tt LIST} a program executed the program instead.)} \\ \hline 
50th & {Modified the reset vector so that pressing {\sc control-reset} caused the Elk Cloner poem to be displayed.} \\ \hline 
55th & {Modified a constant in the diskette calibration code, causing the sound the disk calibration process made during the boot process to change. \cite{Skrenta}} \\ \hline 
60th & {Same as the 55th boot except that a different value was written to the constant in the disk calibration code.} \\ \hline 
65th & {Overwrote the first instruction of the DOS command handler with a jump to the Monitor routine, so that the disk booted into the Monitor.} \\ \hline 
70th & {Same as the 55th boot except that a different value was written to the constant in the disk calibration code.} \\ \hline 
75th-78th & {Unconditionally branched to the first instruction that is executed when a disk booted, leading to four consecutive reboots.} \\ \hline 
79th & {Reset the boot counter.} \\ \hline 
\end{tabular}
\caption{Elk Cloner's manifestation behaviors.  We note that these behaviors are caused by modifying volatile memory.  Therefore, they do not persist after a reboot.}\label{tab:behaviors}
\end{footnotesize}
\end{figure}

The final aspect of Elk Cloner's operation that we discuss is the set of
behaviors that it is most famous for.  Although Elk Cloner manifested itself
using 14 different behaviors (see Figure~\ref{tab:behaviors}), the poem has
been most widely discussed because it was the behavior that was most
unequivocally attributable to something other than a transient failure.  Many
of the ways in which Elk Cloner manifests itself are either subtle and
difficult to observe or appear to be caused by a transient error.  Moreover, a
reboot will always clear up the problem.\footnote{The behavior vanishes because
a reboot causes Elk Cloner's boot counter to increment.}\\
\\
Beginning with the tenth boot, Elk Cloner did something different on every
fifth boot.  On all intervening boot cycles, Elk Cloner returned to DOS without
doing anything.  On the seventy-ninth boot, Elk Cloner reset the boot counter
to zero and the progression started over again.\\

\subsubsection{Obvious Misbehavior}
{\ \\}
Elk Cloner unequivocally made its presence known in precisely one way.  On the
fiftieth boot (and every eightieth boot thereafter), Elk Cloner modified the
reset vector such that it referenced a subroutine that displayed the famous
poem. The reset vector was invoked when the \textsc{reset} and \textsc{control}
keys were pressed simultaneously.  By default, the reset vector pointed to the
DOS warmstart routine; pressing \textsc{control-reset} exited the current
program to a DOS prompt.  Although some programs left the reset vector at its
default value, some programs overwrote the reset vector to perform a
program-specific function (e.g., start a game over again from the beginning).
An Elk Cloner infection would be almost entirely invisible in such a program.
Moreover, without the ability to exit to a DOS prompt, it was impossible for
such a program to spread its infection to other diskettes.\\

\subsubsection{Error-like Behavior}
{\ \\}
Most of the ways in which Elk Cloner manifested itself were easily
mis-attributed to transient software or hardware errors.  For example, on the
tenth boot Elk Cloner overwrote the reset vector with the address of the
Monitor routine.  As a result, when the user pressed \textsc{control-reset} the
Monitor program was entered.  Similarly, on the sixty-fifth boot, Elk Cloner
overwrote the first instruction of the DOS command handler with a jump to the
Monitor routine.  As a result, when the machine booted it enters the Monitor
instead of executing the diskette's \texttt{HELLO} program.  The Monitor was a
low-level interface that enabled the contents of memory to be examined and
modified.  Most notably, the prompt changes from ``\texttt{]}'' to
``\texttt{*}" and DOS commands no longer worked.  Any user who encountered this
would clearly notice the change.  However, if the user rebooted the machine,
the boot counter would increment and the problem would vanish.  \\
\\
On the fifteenth and twenty-fifth boot, Elk Cloner modified the video mode so
that text displayed to the screen was inverted or flashing, respectively.  This
was another case where the effect of Elk Cloner would be obvious to any user,
but it would be quite difficult to attribute to the virus.\\
\\
To invoke DOS commands from Applesoft BASIC programs, the user used the
\texttt{PRINT} command.  The string provided was the desired DOS command,
preceded by a \textsc{control-d}.  \cite{AppleDOS}  The numeric value that was
interpreted as a \textsc{control-d} was stored at address \texttt{\$AAB2}.  On
the thirty-fifth boot, Elk Cloner modified the \textsc{control-d} value.  As a
result, attempts to invoke DOS from an Applesoft program would fail.  The way
that this commonly manifests itself was that instead of executing the
\texttt{HELLO} program on boot, ``\texttt{RUN <PROGRAM>}'' would be printed to
the screen and a DOS prompt would appear.\footnote{Here \texttt{<PROGRAM>} is
the name of the program that would otherwise be executed.}  This was
unmistakably incorrect behavior, but after a reboot the problem vanished.\\
\\
On the fortieth boot, Elk Cloner overwrote the reset vector with an instruction
that executed the reset vector: an infinite loop.  As a result, if
\textsc{control-reset} was pressed the machine would appear to hang, ignoring
all input.  Although this was behavior that the user could not miss, a reboot
resolved the problem.  Elk Cloner exhibited similar behavior on the
seventy-fifth boot.  When Elk Cloner was executed, it unconditionally branched
to \texttt{\$C600}.  This was the memory location of the first instruction
executed when a diskette booted.  This has the effect of rebooting the machine.
Elk Cloner does the same thing on the seventy-sixth through seventy-eighth boot
cycles.  As a result, following the seventy-fifth boot, the machine would
reboot four consecutive times before actually booting the diskette.  This was a
somewhat subtler behavior; the diskette clearly boots more slowly, but only an
attentive user would likely notice the difference.  And because the machine
returns to normal behavior, seemingly of its own accord, it seems likely that
most users would have attributed the change to a transient problem with the
diskette or the drive.  Moreover, given that this was the first virus to
circulate in the wild, users had no reason to suspect tampering if they had
maintained the physical security of their diskettes.\footnote{When Mr.
Skrenta's high school math teacher discovered the virus on one of his
diskettes, he did not suspect a virus but rather accused Mr. Skrenta of
breaking into his office and tampering with his diskettes. \cite{Skrenta}}\\

\subsubsection{Elusive Behavior}
{\ \\}
The remainder of the behaviors that Elk Cloner used to manifest itself were
quite subtle and difficult to detect.  Only a small percentage of users would
likely even notice these behaviors.  For example, on the twentieth boot, Elk
Cloner wrote to the speaker.  This should cause a brief click to be heard from
the speaker.  However, on our machine, the sound was undetectable among all of
the other sounds (e.g., the spinning of the diskette and the movement of the
disk head) that occur during boot.  On the thirtieth boot, Elk Cloner
rearranged the characters that indicate the type of a given file when the
\texttt{CATALOG} command was executed.  Normally, the \texttt{CATALOG} command
identifies Applesoft BASIC files by preceding the filename with an
'\texttt{A}'.  Similarly, binary files are represented by a '\texttt{B}', text
files by a '\texttt{T}' and Integer BASIC files with an '\texttt{I}'.  Elk
Cloner swaps the identifiers for Applesoft BASIC files and binary files and the
identifiers for text files and Integer BASIC files.  For example, following
this change a binary file would be preceded by a '\texttt{A}'.  It is possible
that an attentive user would notice this change.  However, unlike the behaviors
described in the previous section, this behavior would be easily overlooked by
many users.\\
\\
Applesoft BASIC provides a program protection flag whose purpose seems to be an
early (and easily overcome\footnote{An adept user could easily enter the
Monitor and clear the flag.}) attempt at copy-protection.  When the flag was
set, attempts to list or save Applesoft BASIC programs fail.  In this
circumstance, the \texttt{LIST} command causes the program to be executed
instead of displaying the program's listing and the \texttt{SAVE} command
produces a \texttt{PROGRAM TOO LARGE} message.  On the forty-fifth boot, Elk
Cloner sets this flag.  This behavior was not entirely invisible, but it could
easily be overlooked by many users since most common commands (e.g.,
\texttt{CATALOG}, \texttt{RUN}, \texttt{LOAD}, etc.) would continue to function
normally.\\
\\
One of the most subtle behaviors occurs on the fifty-fifth, sixtieth and
seventieth boots.  In these cases, Elk Cloner modifies a constant in the
diskette calibration code.  According to Mr. Skrenta, the changes to this
constant changed the sound the diskette drive made during the calibration
process initiated during the boot process.  \cite{Skrenta}  This change would
only be noticed by the most attentive users.  Most users would either notice no
change or would attribute the change to a transient hardware error since the
machine would appear to operate correctly after booting.\\

\section{Summary and Conclusions}\label{sec:conclude}
Elk Cloner represents an important milestone in computer security.  Its author
had detailed knowledge of Apple hardware and software and used this knowledge
to create a self-replicating program that is now recognized as the first
personal computer virus.  Despite its historical significance, Elk Cloner
received little attention when it was first released.  Elk Cloner was released
in 1982, but its first appearance in the media was in a 1985 Computer
Recreations feature in Scientific American.  \cite{SciAm85}  A Computer
Recreations feature from 1984 devoted to computer viruses made no mention of
Elk Cloner.  \cite{SciAm84}\\
\\
More than three decades after its release, the Apple ][ retains a devoted
following.  Thousands of disk images of Apple ][ programs are currently
available for download.  To partially answer the question of how virulent Elk
Cloner could have been, we downloaded 9,369 distinct Apple II floppy disk
images from BitTorrent and FTP sites.  These images included data disks,
software, commercial games, and pirated games.  One disk image was an exact
match of the Elk Cloner disk image available on Elk Cloner's author's website,
so we excluded it from the analysis.  Of the remaining 9,368 disk images, none
were infected with Elk Cloner.  2,422 (just over 25\%) had no Reloader (i.e.,
were slave diskettes) and contained a version of Apple DOS image that was
potentially capable of being infected by Elk Cloner.  Although more than a
quarter of the images could have been infected by Elk Cloner, none of them
were.\\
\\
We believe that there are several possible reasons that Elk Cloner failed to
cause a greater impact following its release: (1) it spread slowly; (2) it
manifested itself in subtle ways; and (3) it was largely benign.  An Elk Cloner
infection could spread from one diskette to another only when the two diskettes
were used in the same computer.  As a result, the virus could spread only as
rapidly as users physically moved from one computer to the next.  Moreover, it
required a certain set of user behaviors.  While these behaviors may not have
been unusual, any deviation would prevent the virus from spreading.  For
example, a user in the habit of warm-booting her diskettes instead of using the
memory-resident version of DOS to run her programs would avoid infection
entirely.\\
\\
\\
Although Elk Cloner lacked the sophistication of many modern
viruses,\footnote{For example, it makes no attempt to conceal or disguise
itself or avoid removal.} it is an important historical artifact marking the
emergence of viruses in personal computers.  In this paper we have documented
Elk Cloner's behaviors in detail.  When these detailed behaviors are understood
within the context of a historical relationship between computer viruses and
their environments, Elk Cloner sheds new light on today's more sophisticated
and malicious threats.
\end{document}